\documentclass[11pt]{article}

\usepackage{setspace}
\usepackage{algorithm}
\usepackage{algorithmic}
\usepackage{amsfonts}
\usepackage{amsmath}
\usepackage{amssymb}
\usepackage{amsthm}
\usepackage{bm}
\usepackage[justification=centering]{caption}
\usepackage{cite}
\usepackage{color}
\usepackage{epstopdf}
\usepackage{graphicx}

\begin{document}
\newtheorem{definition}{Definition}[]
\renewcommand{\algorithmicrequire}{\textbf{Input:}}
\renewcommand{\algorithmicensure}{\textbf{Output:}}
\title{A Security-assured Accuracy-maximised Privacy Preserving Collaborative Filtering Recommendation Algorithm}
\date{}

\author{
Zhigang Lu\\
       The University of Adelaide\\
       Adelaide, Australia\\
       zhigang.lu@adelaide.edu.au
\and
Hong Shen\\
			The University of Adelaide\\
			Adelaide, Australia\\
			hong.shen@adelaide.edu.au}

\maketitle
~\\
\begin{abstract}
The neighbourhood-based Collaborative Filtering is a widely used method in recommender systems. However, the risks of revealing customers' privacy during the process of filtering have attracted noticeable public concern recently. Specifically, $k$NN attack discloses the target user's sensitive information by creating $k$ fake nearest neighbours by non-sensitive information. Among the current solutions against $k$NN attack, the probabilistic methods showed a powerful privacy preserving effect. However, the existing probabilistic methods neither guarantee enough prediction accuracy due to the global randomness, nor provide assured security enforcement against $k$NN attack. To overcome the problems of current probabilistic methods, we propose a novel approach, Partitioned Probabilistic Neighbour Selection, to ensure a required security guarantee while achieving the optimal prediction accuracy against $k$NN attack. In this paper, we define the sum of $k$ neighbours' similarity as the accuracy metric $\alpha$, the number of user partitions, across which we select the $k$ neighbours, as the security metric $\beta$. Differing from the present methods that globally selected neighbours, our method selects neighbours from each group with exponential differential privacy to decrease the magnitude of noise. Theoretical and experimental analysis show that to achieve the same security guarantee against $k$NN attack, our approach ensures the optimal prediction accuracy.
\end{abstract}
~\\


{\bf Topics} Differential Privacy, Privacy Metrics, Privacy in Recommender Systems
~\\

\section{Introduction}
\label{INTRO}

$Recommender$ $systems$ predict customers' potential preferences by aggregating history data and customers' interests. Recently, the increasing importance of recommender systems in various Internet applications should be noticed. For example, Amazon has been receiving benefits for a decade from the recommender systems by providing personal recommendation to their customers, and Netflix posted a one million U.S. dollars award for improving their recommender system to make their business more profitable \cite{SCHAFER1999, EKSTRAND2011, KABBUR2013}. Currently, in recommender systems, Collaborative Filtering (CF) is a famous technology with three main popular algorithms \cite{LIU2011}, i.e., neighbourhood-based methods \cite{EKSTRAND2011}, association rules based prediction \cite{HERLOCKER2002}, and matrix factorisation \cite{KOREN2009}. Among these algorithms, neighbourhood-based methods are the most widely used in the industry because of its easy implementation and high prediction accuracy.

One of the most popular neighbourhood-based CF method is $k$ Nearest Neighbour ($k$NN) method which provides recommendations by aggregating the opinions of a user's $k$ nearest neighbours \cite{ADOMAVICIUS2005}. Although $k$NN method efficiently presents good recommendation performance of accuracy, the risk of customers' privacy disclosure during the process of filtering is a growing concern, e.g., the $k$NN attack \cite{CALANDRINO2011} which exploits the property that the users are more similar when sharing same ratings on corresponding non-sensitive items to reveal user's sensitive information. Thus proposing an efficient privacy preserving neighbourhood-based CF algorithm against $k$NN attack, which obtains trade-off between the system security and recommendation accuracy, has been a natural research problem.

The literature in CF recommender systems has shown several approaches to preserve customers' privacy. Generally, cryptographic methods, obfuscation, perturbation, randomised methods (including naive probabilistic methods and differential privacy methods) are applied \cite{ZHU2014}. Among them, cryptographic methods \cite{ERKIN2010, NIKOLAENKO2013} provide the most reliable security but the unnecessary computational cost cannot be ignored. Obfuscation methods \cite{PARAMESWARAN2007, WEINSBERG2012} and Perturbation methods \cite{BASU2012, BILGE2012} introduce designed random noise into the original matrix to preserve customers' sensitive information; however the magnitude of noise is hard to calibrate in these two types of methods \cite{DWORK2006B, ZHU2014}. The naive probabilistic method \cite{ADAMOPOULOS2014} provides a similarity based weighted neighbour selection for the $k$ neighbours. Similar to perturbation, McSherry et al. \cite{MCSHERRY2009} presented a naive differential privacy method which adds calibrated noise into the covariance (similarity between users/items) matrix. Similar to the naive probabilistic neighbour selection \cite{ADAMOPOULOS2014}, Zhu et al. \cite{ZHU2014} proposed a Private Neighbour CF to preserve privacy against $k$NN attack by introducing differential privacy in selecting the $k$ nearest neighbours randomly, then adding Laplace differential noise into covariance matrix. Although the methods in \cite{MCSHERRY2009, ZHU2014, ADAMOPOULOS2014} successfully preserve users' privacy against $k$NN attack, the low prediction accuracy due to the global randomness should be remarked. Moreover, as privacy preserving CF recommendation algorithms, none of the existing randomised methods provide an assured security enforcement before the process of filtering.

\textbf{Contributions.} In this paper, to overcome the problems of unsatisfactory prediction accuracy and unassured security guarantee in the existing probabilistic approaches against $k$NN attack, we propose a novel method, Partitioned Probabilistic Neighbour Selection. The main contributions of this paper are:
\begin{itemize}
\item We define performance metrics clearly in both prediction accuracy and system security to theoretically analyse the performance of privacy preserving CF method. Specifically, we define the sum of $k$ neighbours' similarity as the accuracy metric $\alpha$, the number of user partitions, across which we select the $k$ neighbours, as the security metric $\beta$.
\item We propose a novel differential privacy preserving method, Partitioned Probabilistic Neighbour Selection (PPNS), which achieves the optimal prediction accuracy $\alpha$ with a given desired system security $\beta$ among all of the existing developments of randomised neighbourhood-based CF recommendation algorithms.
\item We show that, compared with the related methods, the proposed PPNS method performs consistently well across various experimental settings. For example, we compare the accuracy performance on different datasets; we design the experiments on both user-based and item-based neighbourhood-based CF; we examine the accuracy performance in the scenario with and without $k$NN attack.
\end{itemize}
\textbf{Organisation.} The rest of this paper is organised as follows: Firstly, in Section \ref{RW}, we discuss both the advantages and disadvantages in the existing privacy preserving methods on CF recommender systems. Then we introduce the relevant preliminaries in this paper in Section \ref{PRE}. Afterwards, we present a classic attacking against neighbourhood-based CF recommender systems in Section \ref{ATK}. Next, we propose a novel differential privacy recommendation approach, Partitioned Probabilistic Neighbour Selection in Section \ref{SCH}. In Section \ref{PE}, the experimental analysis of our approach on the performance of both recommendation accuracy and security are provided. Finally, we conclude with a summary in Section \ref{CON}.

\section{Related Work}
\label{RW}

A noticeable number of literature has been published on privacy risks to preserve customers' private data in recommender systems. In this section, we briefly discuss some of the research literature in privacy preserving CF recommender systems.

\subsection{Traditional Privacy Preserving CF Recommendation}
Amount of traditional privacy preserving methods have been developed in CF recommender systems \cite{ZHU2014}, including cryptographic \cite{ERKIN2010, NIKOLAENKO2013}, obfuscation \cite{PARAMESWARAN2007, WEINSBERG2012}, perturbation \cite{BASU2012, BILGE2012} and probabilistic methods \cite{ADAMOPOULOS2014}. Erkin et al. \cite{ERKIN2010} applied homomorphic encryption and secure multi-party computation in privacy preserving recommender systems, which allows users to jointly compute their data to receive recommendation without sharing the true data with other parties. Nikolaenko et al. \cite{NIKOLAENKO2013} combined a famous recommendation technique, matrix factorization, and a cryptographic method, garbled circuits, to provide recommendations without learning the real user ratings in database. The Cryptographic methods provide the highest guarantee for both prediction security ans system security by introducing encryption rather than adding noise to the original record. Unfortunately, unnecessary computational cost impacts its application in industry \cite{ZHU2014}. Obfuscation and perturbation are two similar data processing methods. In particular, obfuscation methods aggregate a number of random noises with real users rating to preserve user's sensitive information. Parameswaran et al. \cite{PARAMESWARAN2007} proposed an obfuscation framework which exchanges the sets of similar items before submitting the user data to CF server. Weinsberg et al. \cite{WEINSBERG2012} introduced extra reasonable ratings into user's profile against inferring user's sensitive information. Perturbation methods modify the user's original ratings by a selected probability distribution before using these ratings. Particularly, Bilge et al. \cite{BILGE2012} added uniform distribution noise to the real ratings before the utilisation of user's rating in prediction process. While, Basu et al. \cite{BASU2012} regarded the deviation between two items as the adding noise. Both perturbation and obfuscation obtain good trade-off between prediction accuracy and system security due to the tiny data perturbation, but the magnitude of noise or the percentage of replaced ratings are not easy to be calibrated \cite{DWORK2006B, ZHU2014}. The naive probabilistic method \cite{ADAMOPOULOS2014} applied weighted sampling in the process of neighbour selection which preserves users' privacy against $k$NN attack successfully, because of the perturbation of the final neighbour set; however, it cannot guarantee enough prediction accuracy due to the global randomness. Moreover, these traditional privacy preserving CF methods are unable to measure privacy levels against $k$NN attack, thus impairing the credibility of the final recommendation result.

\subsection{Differential Privacy CF Recommendation}
As a well-known privacy definition, the differential privacy mechanism \cite{DWORK2006} has been applied in the research of privacy preserving recommender systems. For example, McSherry et al. \cite{MCSHERRY2009} provided the first differential privacy neighbourhood-based CF recommendation algorithm. Actually, the naive differential privacy protects the neighbourhood-based CF recommender systems against $k$NN attack successfully, as they added Laplace noise into the covariance matrix globally, so that the output neighbour set is no longer the original $k$ neighbours ($k$ nearest candidates). However, the prediction accuracy of their recommendation algorithm is decreased significantly due to the introduction of global noise.

Another development of differential privacy neighbourhood-based CF method, Private Neighbour CF (PNCF), is proposed by \cite{ZHU2014} which inspires our work. They theoretically fixed the low prediction accuracy problem of naive probabilistic neighbour selection \cite{ADAMOPOULOS2014} by a truncated parameter $\lambda$. As a differential privacy method, the selection weight in PNCF method is measured by the following equation:
\begin{equation}
\label{OMEGA}
\omega_i=\exp(\frac{\epsilon}{4k\times RS}q_a(U(u_a),u_i)),
\end{equation}
where $\epsilon$ is differential privacy budge, $q$ is the score function, $RS$ is the Recommendation-Aware Sensitivity of score function $q$ for any user pair $u_i$ and $u_j$, and $U(u_a)$ is the set of user $u_a$'s candidate list. For a user $u_a$, the score function $q$ and its Recommendation-Aware Sensitivity are defined as:
\begin{equation}
\label{SelectionFunc}
q_a(U(u_a),u_i)=sim(a,i),
\end{equation}
\begin{equation}
\label{RS}
RS=\max\left\{ \underset{s\in S_{ij}}{\max}\left(\frac{r_{i,s}\cdot r_{j,s}}{\left\Vert r_{i}'\right\Vert \left\Vert r_{j}'\right\Vert }\right),\underset{s\in S_{ij}}{\max}\left(\frac{r_{i,s}\cdot r_{j,s}\left(\left\Vert r_{i}\right\Vert \left\Vert r_{j}\right\Vert -\left\Vert r_{i}'\right\Vert \left\Vert r_{j}'\right\Vert \right)}{\left\Vert r_{i}\right\Vert \left\Vert r_{j}\right\Vert \left\Vert r_{i}'\right\Vert \left\Vert r_{j}'\right\Vert }\right)\right\},
\end{equation}
where $r_{i,s}$ is user $u_i$'s rating on item $t_s$, $sim(a,i)$ is the similarity between user $u_a$ and $u_i$, $r_i$ is user $u_i$'s average rating on every item, $S_{ij}$ is the set of all items co-rated by both users $i$ and
$j$, i.e., $S_{ij}=\{s\in S|r_{i,s}\neq \varnothing \ \&\
r_{j,s}\neq \varnothing \}$.

Then, the PNCF method selects the $k$ neighbours which include the candidates whose similarity is greater than $(sim_k+\lambda)$ and randomised candidates whose similarity is between $(sim_k+\lambda)$ and $(sim_k-\lambda)$, where $sim_k$ denotes the similarity of the $k$th candidate of a target user. Zhu et al. \cite{ZHU2014} provided an equation to calculate the value of $\lambda$, i.e. $\lambda=\min(sim_k,\frac{4k\cdot RS}{\epsilon}\ln\frac{k(n-k)}{\rho})$, where $\rho$ is a constant, $0<\rho<1$. Once having the $k$ neighbours set, Zhu et al. \cite{ZHU2014} added Laplace differential noise in the final $k$ neighbour's similarity matrix to perturb the final prediction. Their experimental results showed better prediction performance than \cite{MCSHERRY2009}.

We observe that PNCF \cite{ZHU2014} has two weaknesses. Firstly, it unnecessarily adds random noise in the process of filtering twice (one at neighbour selection stage, another at rating prediction stage), the extra randomness will decrease the prediction accuracy significantly. Secondly, the value of $\lambda$ may not be achievable. This is because the computation of $\lambda$ results in a good theoretical recommendation accuracy, but does not yield a good experimental prediction accuracy on the given test data sets in \cite{ZHU2014} against $k$NN attack. So the PNCF method \cite{ZHU2014} will actually be a method of naive Probabilistic Neighbour Selection \cite{ADAMOPOULOS2014} and cannot guarantee enough recommendation accuracy.

In conclusion, compared with cryptographic, obfuscation and perturbation privacy preserving methods, the probabilistic methods are more efficient. The existing Probabilistic solutions \cite{MCSHERRY2009,ADAMOPOULOS2014,ZHU2014} on privacy preserving neighbourhood-based CF recommender systems applied different randomised strategies to improve the prediction accuracy, while ensure the security against $k$NN attack by selecting the $k$ neighbours across a target user's partial/entire candidate list. However, they failed to guarantee enough prediction accuracy due to the introduction of global noise. Additionally, as privacy preserving CF recommendation algorithms, none of the existing randomised methods provide an assured security enforcement before the process of CF recommendation against $k$NN attack. Therefore, in this paper, we aim to propose a randomised privacy preserving neighbourhood-based CF recommendation algorithm which guarantees an assured security firstly, then achieves the optimal prediction accuracy with the assured security guarantee.

\section{Preliminaries}
\label{PRE}
In this section, we introduce the foundational concepts related with this paper in collaborative filtering, differential privacy, and Wallenius' non-central hypergeometric distribution.

\subsection{$k$ Nearest Neighbour Collaborative Filtering}
$k$ Nearest Neighbour collaborative filtering is the most popular recommendation method in neighbourhood-based CF recommender systems, which predicts customer's potential preferences by aggregating the opinions of the $k$ most similar neighbours \cite{ADOMAVICIUS2005}.

Neighbour Selection and Rating Prediction are two main steps in neighbourhood-based CF \cite{ZHU2014}. At the Neighbour Selection stage, $k$ nearest candidates are selected from the target user $u_a$'s candidate list $\mathcal{S}_a$, where similarities between $u_a$ and any other users are calculated by similarity measurement metric. There are two famous similarity measurement metrics: the Pearson correlation coefficient and Cosine-based similarity \cite{ADOMAVICIUS2005}. In this paper, we use the Cosine-based similarity \cite{RAJARAMAN2011} as the similarity measurement metric because of its lower complexity.
\begin{equation}
\label{COS}
sim(i,j)=\frac{\sum_{s\in S_{ij}}r_{i,s}r_{j,s}}{\sqrt{\sum_{s\in S_{i}}r_{i,s}^2}\sqrt{\sum_{s\in S_{j}}r_{j,s}^2}},
\end{equation}
where $sim(i,j)$ is the similarity between user $u_i$ and $u_j$, $r_{i,s}$ is user $u_i$'s rating on item $t_s$, $r_{i,s}\in\mathcal{R}$, $\mathcal{R}$ is the user-item rating dataset, $\bar{r_i}$ is user $u_i$'s average rating on every item, $S_{ij}$ is the set of all items co-rated by both user $u_i$ and $u_j$, i.e., $S_{ij}=\{s\in S|r_{i,s}\neq \varnothing \ \&\ r_{j,s}\neq \varnothing \}$, $S_{i}$ is the set of all items rated by user $u_{i}$, i.e., $S_{i}=\{s\in S|r_{i,s}\neq \varnothing \}$.

At the stage of Rating Prediction, to predict the potential rating $\hat{r}_{ax}$ of user $u_a$ on item $t_x$, all ratings on $t_x$ of the $k$ selected users (which are called neighbours) will be aggregated. For example, for user-based methods, the prediction of $\hat r_{ax}$ is shown as below:
\begin{equation}
\label{PRED}
\hat{r}_{ax}=\frac{\sum_{u_i\in
N_k(u_a)}sim(a,i)r_{i,x}}{\sum_{u_i\in N_k(u_a)}|sim(a,i)|},
\end{equation}
where, $N_k(u_a)$ is a sorted set which contains user $u_a$'s $k$ nearest neighbours, $N_k(u_a)$ is sorted by similarity in a descending order, $sim(a,i)$ is the $i$th neighbour of $u_a$ in $N_k(u_a)$.

\subsection{Differential Privacy}
Informally, differential privacy \cite{DWORK2006} is a scheme that minimises the sensitivity of output for a given statistical operation on two different (differentiated in one record to protect) datasets. Specifically, differential privacy guarantees whether one specific record appears or does not appear in a database, the privacy mechanism will shield the specific record to the adversary. The strategy of differential privacy is adding a random calibrated noise to the result of a query function on the database. We say two datasets $X$ and $X^{\prime}$ are neighbouring dataset, if they differ in only one record at most. A formal definition of Differential Privacy is shown as follows:
\begin{definition}[$\epsilon$-Differential Privacy \cite{DWORK2006}]
A randomised mechanism $\mathcal{T}$ is $\epsilon$-differential privacy if for all neighbouring datasets $X$ and $X^{\prime}$, and for all outcome sets $S\subseteq Range(T)$, $\mathcal{T}$ satisfies: $\Pr[\mathcal{T}(X)\in S]\leq exp(\epsilon)\cdot\Pr[\mathcal{T}(X^{\prime})\in S]$, where $\epsilon$ is a privacy budget. 
\end{definition}
\begin{definition}[Exponential Differential Privacy Mechanism \cite{MCSHERRY2007}]
Given a score function of a database $X$, $q(X,x)$, which reflects the score of query respond $x$. The exponential mechanism $\mathcal{T}$ provides $\epsilon$-differential privacy, if $\mathcal{T}(X)$ = \{the probability of a query respond $x$ $\propto$ $\exp({\frac{\epsilon\cdot q(X,x)}{2\Delta q}})$\}, where $\Delta q=\max|q(X,x)-q(X^{\prime},x)|$, denotes the sensitivity of $q$.
\end{definition}

\subsection{Wallenius' Non-central Hypergeometric Distribution}
\label{WNHD}
Briefly, Wallenius' Non-central Hypergeometric Distribution is a distribution of weighted sampling without replacement \cite{WALLENIUS1963}. We assume there are $c$ categories in the population, category $i$ contains $m_i$ individuals. All the individuals in category $i$ have the same weight $\omega_i$. The probability of an individual is sampled at a given draw is proportional to its weight $\omega_i$.

In this paper, we use the following properties of Wallenius' Non-central Hypergeometric Distribution to find the optimal prediction accuracy neighbour selection with a given security guarantee against $k$NN attack. \cite{MANLY1974} gave the approximated solution to the mean $\bm{\mu}=(\mu_{1},\mu_{2},\ldots,\mu_{c})$ of $\bm{x}=(x_1,x_2,\ldots,x_c)$, where $x_i$ denotes the number of individuals sampled from category $i$ by Wallenius' Non-central Hypergeometric Distribution, $\sum_{i=1}^{c}{x_i}=\sum_{i=1}^{c}{\mu_i}=k$:
\begin{equation}
\label{approx}
\left(1-\frac{\mu_1^\star}{m_1}\right)^{1/\omega_1}=\left(1-\frac{\mu_2^\star}{m_2}\right)^{1/\omega_2}=\ldots=\left(1-\frac{\mu_c^\star}{m_c}\right)^{1/\omega_c},
\end{equation}
where $\sum_{i=1}^{c}{\mu_i^\star}=k$, $\forall i\in C: 0\leq \mu_i^\star \leq m_i.$

The solution $\bm{\mu}^\star=(\mu_{1}^\star,\mu_{2}^\star,\ldots,\mu_{c}^\star)$ is an approximation to the mean $\bm \mu$. Fog \cite{FOG2008} stated the following properties of Equation \eqref{approx}: firstly, the solution $\bm{\mu}^*$ is valid under the conditions that $\forall i \in C: m_i>0$ and $\omega_i>0$. Secondly, the mean given by Equation \eqref{approx} is a good approximation in most cases. Thirdly, Equation \eqref{approx} is exact when all $\omega_i$ are equal.

\section{A Privacy Attacking on CF Recommender Systems}
\label{ATK}
In this section, we introduce a classic neighbourhood-based CF attacking, $k$ Nearest Neighbour ($k$NN) attack. Calandrino et al. \cite{CALANDRINO2011} presented a user-based attacking, \textit{k Nearest Neighbour} ($k$NN) attack, against the $k$NN CF recommendation algorithm. Simply, $k$NN attack exploits the property that the users are more similar when sharing same ratings on corresponding items to reveal user's private data.

We suppose that an attacker's background knowledge consists of both the recommendation algorithm ($k$NN CF recommendation) and its parameter $k$. Furthermore, a target user $u_a$'s partial non-sensitive history ratings, i.e., the ratings on $m$ items that $u_a$ voted, are known to the attacker.

The aim of $k$NN attack is to disclose $u_a$'s sensitive transactions that the attacker does not yet know about. To achieve this goal, the attacker firstly registers $k$ fake users in a $k$NN recommender system who only vote on $u_a$'s $m$ non-sensitive items with same ratings of $u_a$. With a high probability, each fake user's $k$ nearest neighbours set $N_k(\text{fake user})$ will include the other $k-1$ fake users and the target user $u_a$. Because the target user $u_a$ is the only neighbour who has ratings on the items which are not rated by the fake users, to provide recommendations on these items to the fake users, the recommender system has to give $u_a$'s rating to the fake users directly. Obviously, the fake users learn the target user $u_a$'s whole rating list successfully with $k$NN attack.

\section{Privacy Preservation by Partitioned Probabilistic Neighbour Selection}
\label{SCH}
In this section, we firstly present the motivations and the goal of this paper. Then we provide two performance metrics on privacy preserving neighbourhood-based CF recommender systems against $k$NN attack. Finally we propose our Partitioned Probabilistic Neighbour Selection algorithm based on our motivations and goal.

\subsection{Motivation}
Current research \cite{MCSHERRY2009,ZHU2014,ADAMOPOULOS2014} on privacy preserving neighbourhood-based CF recommender systems applied different randomised strategies to improve the prediction accuracy, while ensure the security against $k$NN attack by selecting the $k$ neighbours across a target user's partial/entire candidate list. Among these randomised strategies, differential privacy is a better privacy preserving mechanism as it provides calibrated magnitude of noise.

Actually, since the information collected by recommender systems is always the customers' personal data \cite{CALANDRINO2011}, preserving the users' sensitive information should be the kernel issue of recommender systems. But none of the existing privacy preserving neighbourhood-based CF recommendation algorithms ensure a successful security-assured privacy preservation against $k$NN attack before the process of CF recommendation. So in this paper, we present a security metric to measure the level of system security.

In addition, the prediction accuracy should also be considered carefully with the guarantee of assured security, otherwise, the recommender systems would be useless to the non-malicious users who are the majority of customers. However, because of the introduction of global noise, the current randomised methods cannot guarantee the prediction accuracy either. To provide enough prediction utility, we have to decrease the noise as much as possible. Since there is no need to add noise into both the stage of neighbour selection and rating prediction, we may simply add Laplace noise \cite{DWORK2006} to the final prediction rating after a regular $k$NN CF. Unfortunately, as Sarathy et al. \cite{SARATHY2011} reported the security risk about the Laplace mechanism for numeric data, the above idea should be rejected. So we focus on adding noise at the stage of neighbour selection. Instead of global neighbour selection, we partition the order candidate list, so that we can control magnitude of noise inside each partition.

Therefore, in this paper, we aim to propose a partitioned probabilistic (differential privacy) neighbour selection method, which guarantees an assured security, then achieves the maximum prediction accuracy with the assured security against $k$NN attack, without any perturbations in the process of rating prediction.

\subsection{Performance Metrics}
\subsubsection{Accuracy}
Naturally, in any neighbourhood-based CF recommender systems, aggregating the ratings of more similar users yields more reliable prediction. Therefore, we define the accuracy performance metric $\alpha$ as the similarity sum of the $k$ neighbours of a target user $u_a$. Obviously, the greatest value of $\alpha$ would be the similarity sum of the $k$ nearest candidates of a target user.

It is simple to compute $\alpha$ in the deterministic neighbourhood-based CF algorithms, e.g. $k$NN CF recommendation algorithm, because the $k$ neighbours selected by the deterministic algorithms are determined. So in the case of deterministic algorithms, we compute $\alpha$ by the following equation,
\begin{equation}
\label{DETA}
\alpha=\sum_{i=1}^k{sim(a, neighbour_i)}.
\end{equation}

While, in the randomised neighbourhood-based CF algorithms, because of the randomisation, we should calculate the value of $\alpha$ as the expected similarity sum of the $k$ neighbours by
\begin{equation}
\label{RANDA}
\alpha=\mathbb{E}(\sum_{i=1}^k{sim(a,neighbour_i)}).
\end{equation}

However, it is difficult to compute Equation \eqref{RANDA} directly, as we need to find all the possible $k$-neighbour combinations and their corresponding probabilities. So we give another way to compute the expectation in Equation \eqref{RANDA}, shown in blow:
\begin{equation}
\label{EXPEC}
\begin{array}{ccl}
\alpha & = & \mathbb{E}(\sum_{i=1}^{k}{sim(a,neighbour_{i})})\\
 & = & \sum_{i=1}^{n}sim(a,user_{i})\mathbb{E}(x_{i})\\
 & = & \sum_{i=1}^{n}{sim(a,user_{i})\mu_{i}},
\end{array}
\end{equation}
where $\sum_{i=1}^n{\mu_{i}}=k$, $\mu_{i}\in [0,1)$. Section \ref{WNHD} introduced the definition of $x_i$ and $\mu_i$.

Actually, when $user_i$ is selected as a neighbour of the target user $u_a$, $\mu_{i}=1$, while when $user_i$ is not a neighbour of $u_a$, $\mu_{i}=0$. Namely, in this paper, deterministic algorithms (Equation \eqref{DETA}) is a special case of randomised algorithms (Equation \eqref{EXPEC}). Therefore, we compute the accuracy metric $\alpha$ by the following equation in both deterministic and randomised neighbourhood-based CF recommendation algorithms:
\begin{equation}
\label{ACCURACY}
\left\{ \begin{array}{l}
\alpha=\sum_{i=1}^{n}{sim(a,user_{i})\mu_{i}},\\
k=\sum_{i=1}^{n}{\mu_{i}}.
\end{array}\right.
\end{equation}

\subsubsection{Security}
According to the property of $k$NN attack, the purpose of a privacy preserving neighbourhood-based CF recommendation algorithm is to avoid the target user being the only real user in the final $k$ neighbours set. Thus, the existing probabilistic privacy preserving solutions select the $k$ neighbours across the partial/entire candidate list. It is obvious that the number of candidates who may be selected into the $k$ neighbours set decides the success probability of $k$NN attack (we call these candidates as potential neighbours). Namely, the more potential neighbours result in the less probability that the target user is the only real user in the final $k$ neighbours set. On the other side, the attacker needs to create enough fake users to cover the potential neighbours set, so that the target user can be the only real user. That is to say, the more potential neighbours yield the higher attacking cost. In conclusion, in this paper, because we partition the candidate list by the given $k$, we define the number of user partitions, across which we select the $k$ neighbours, as the security metric $\beta$.

\subsection{Partitioned Probabilistic Neighbour Selection Scheme}
To achieve our goal, we will firstly provide the objective function with its constraints based on the discussions on both two performance metrics. Then, we propose the security-assured accuracy-maximised privacy preserving recommendation method by solving the objective function according to its constraints.

According to the security metric $\beta$ and the properties of $k$NN attack, we partition the entire candidate list of a user by the given $k$, i.e., the size of each partition (group) is $k$. Before providing the objective function, we introduce some variables in advance. We use $f_{\beta}(i)$ to denote the number of neighbours selected (weighted sampling with exponential differential privacy) from partition No. $i$ with the given security metric $\beta$, $i \in [1,\beta]$. Additionally, $\alpha_i$ denotes the prediction accuracy of partition No. $i$ against $k$NN attack. Therefore, we have a general equation for $\alpha$,
\begin{equation}
\label{ALPHA}
\alpha = \sum_{i=1}^{\beta}{\alpha_i}.
\end{equation}

To solve the Equation \eqref{ALPHA} for the optimal $\alpha$ with the given security metric $\beta$ against $k$NN attack, we select one random fake user as the user who receives the system recommendation. We suppose the candidate list of the fixed fake user has already in a descending order of similarity. Figure \ref{fig:1kNN} shows the fixed fake user's candidate list, where $N_i$ denotes to the user set in partition $i$, $i \in [2, \beta]$, $u_a$ is the attacker's target user.
\begin{figure}[!h]
\centering
\begin{tabular}{|c|c|c|c|c|c|}
\hline 
Partition Number & 1 & 2 & $\cdots$ & $\beta-1$ & $\beta$\tabularnewline
\hline 
Partition Content & Fake users + $u_a$ & $N_{2}$ & $\cdots$ & $N_{\beta-1}$ & $N_{\beta}$\tabularnewline
\hline 
\end{tabular}
\caption{Candidate list against $k$NN attack}
\label{fig:1kNN}
\end{figure}

According to formulas \eqref{ACCURACY} and Figure \ref{fig:1kNN}, we have
\begin{equation}
\label{ALPHAi}
\alpha_i = \sum_{j=1}^{k}{sim_{j,N_i}\mu_{j,N_i}},
\end{equation}
where $sim_{j,N_i}$ denotes the similarity between $j$th candidate in partition No. $i$ and the fixed fake user, $\mu_{j,N_i}$ denotes the corresponding mean $\mu$, $in \in [1,\beta]$. Moreover, because we aim to select $f_{\beta}(i)$ neighbours from partition No. $i$, $\sum_{j=1}^{k}{\mu_{j,N_i}}=f_{\beta}(i)$.

Combining Equation \eqref{ALPHA} and Equation \eqref{ALPHAi}, we have 
\begin{equation}
\alpha = \sum_{i=1}^{\beta}{\sum_{j=1}^{k}{sim_{j,N_i}\mu_{j,N_i}}}.
\end{equation}
Since the similarity between the candidates in partition No. 1 and the fixed fake users is absolutely one, we rewrite the above equation as
\begin{equation}
\label{OBJ}
\alpha = f_{\beta}(1) + \sum_{i=2}^{\beta}{\sum_{j=1}^{k}{sim_{j,N_i}\mu_{j,N_i}}}.
\end{equation}
Obviously, the Equation \eqref{OBJ} is our objective function against $k$NN attack.

Now we give the constraints of Equation \eqref{OBJ}. Since we need to select the $k$ neighbours across the top $\beta$ partitions, we should select at least one neighbour from partition No. $\beta$, i.e., $f_{\beta}(\beta) = \sum_{i=1}^k{\mu_{i,N_\beta}} \geq 1$. As the candidate list is in a descending order of similarity, and we select one neighbour from the partition No. $\beta$, to cover all the top $\beta$ partitions, the attacker needs to create at least $\beta k$ fake users, no matter how many neighbours are selected from the partition No. $i$, $i \in [1,\beta-1]$. So we can select zero neighbour from the partition No. $i$, $i \in [1,\beta-1]$. In addition, because $f_{\beta}(\beta) \geq 1$ and $\sum_{i=1}^{\beta}f_{\beta}(i)=k$, $f_{\beta}(i) \leq k-1$ for $i \in [1, \beta-1]$. Recalling the other constraints we presented previously, we have the final objective function with constraints as follow:
\begin{equation}
\label{FINAL}
\begin{array}{ll}
\text{maximise} & \alpha = f_{\beta}(1) + \sum_{i=2}^{\beta}{\sum_{j=1}^{k}{sim_{j,N_i}\mu_{j,N_i}}}\\
\text{subject to} & \sum_{j=1}^{k}\mu_{j,N_i}=f_{\beta}(i)\\
 & \sum_{i=1}^{\beta}f_{\beta}(i)=k\\
 & f_{\beta}(i)\in\left\{ \begin{array}{ll}
[1,k], & i=\beta\\{}
[0,k-1], & i \in [1,\beta)
\end{array}\right.
\end{array}
\end{equation}

Then, we solve Linear Programming \eqref{FINAL} as a Knapsack Problem with the property of Equation \eqref{approx}. The solution, that is the partitioned probabilistic neighbour selection method which guarantees the optimal expectation of prediction accuracy $\alpha$ with a given security metric $\beta$ against $k$NN attack is:
\begin{equation}
\label{MaxSelection}
f_{\beta}(i)=\left\{ \begin{array}{cl}
k-1, & i=1\\
1, & i=\beta\\
0, & i \in (1,\beta)
\end{array}\right..
\end{equation}
Note that because $\forall$ $\beta \geq 1$, the candidate list of any user is in a descending order of similarity, formula \eqref{MaxSelection} will always be the optimal solution to Linear Programming \eqref{FINAL} for any $\beta \geq 1$.

Algorithm \ref{ALG} demonstrates the Partitioned Probabilistic Neighbour Selection (PPNS) method. From line 1 to line 5, we compute the necessary parameters by Equation \eqref{COS}, \eqref{RS}, \eqref{SelectionFunc} and \eqref{OMEGA}. We select the $k$ neighbours from each partition with exponential differential privacy by Partitioned Probabilistic Neighbour Selection (Equation \eqref{MaxSelection}) in line 6. Next, once we have the $k$ neighbours of target user $u_a$, we compute the prediction rating of $u_a$ on a item $r_x$, $r_{ax}$, by Equation \eqref{PRED} in line 7. Finally, we return the neighbour set $N_{k}(u_a)$ and the prediction rating $r_{ax}$.
\begin{algorithm}[ht]
\caption{Partitioned Probabilistic Neighbour Selection.}
\label{ALG}
\begin{algorithmic}[1]
\REQUIRE ~~\\
Original user-item rating set, $\mathcal{R}$;\\
Target user, $u_a$ and prediction item, $t_x$;\\
Number of neighbours, $k$;\\
Differential privacy parameter, $\epsilon$;\\
Security metric, $\beta$.
\ENSURE ~~\\
Target user $u_a$'s $k$-neighbour set, $N_{k}(u_a)$;\\
Prediction rating of $u_a$ on $t_x$, $r_{ax}$.
\STATE Compute the similarity array for target user $u_a$, $\mathcal{S}_a$;
\STATE Sort $\mathcal{S}_a$ in descending order, $\mathcal{S}_a^\prime$;
\STATE Compute exponential differential privacy sensitivity, $RS$;
\STATE Compute each user $u_i$'s selection weight, $\omega_i$;
\STATE Partition the sorted $\mathcal{S}_a^\prime$ by $k$;
\STATE Select $k$ neighbours from top $\beta$ partitions;
\STATE Compute $r_{ax}$ by $N_{k}(u_a)$;
\RETURN $N_{k}(u_a)$, $r_{ax}$;
\end{algorithmic}
\end{algorithm}

\section{Experimental Evaluation}
\label{PE}
In this section, we use the real-world datasets to evaluate the performance on both accuracy and security of our Partitioned Probabilistic Neighbour Selection method. We begin by the description of the datasets, then introduce the evaluation metric, finally perform a comparative analysis of our method and some existing privacy preserving neighbourhood-based CF recommendation algorithms.

\subsection{Data set and Evaluation Metric}
In the experiments, we use two real-world datasets, MovieLens dataset\footnote{http://www.grouplens.org/datasets/movielens/} and Douban\footnote{http://www.douban.com} (one of the largest rating websites in China) film dataset\footnote{https://www.cse.cuhk.edu.hk/irwin.king/pub/data/douban}. The MovieLens dataset consists of 100,000 ratings (1-5 integral stars) from 943 users on 1682 films, where each user has voted more than 20 films, and each film received 20$-$250 users' rating. The Douban film dataset contains 16,830,839 ratings (1-5 integral starts) from 129,490 unique users on 58,541 unique films \cite{MA2011}.

We use a famous measurement metric, Mean Absolute Error (MAE) \cite{WILLMOTT2005, ZHU2014}, to measure the recommendation accuracy, in the experiments:
\begin{equation}
\label{MAE}
MAE=\frac{1}{UI}\sum_{i=1}^{U}{\sum_{j=1}^{I}{|r_{ij}-\hat{r}_{ij}|}},
\end{equation}
where $r_{ij}$ is the real rating of user $u_i$ on item $t_j$, and $\hat{r}_{ij}$ is the corresponding predicted rating from recommendation algorithms, $U$ and $I$ denote the number of users and items in the experiments. Specifically, in user-based experiments, we compute the $MAE$ of ratings from 200 random users ($U=200$) on all the items ($I=1682$ or $I=58,541$) in the two datasets, while, in item-based experiments, we compute the $MAE$ of ratings on 200 random items ($I=200$) from all the users ($U=943$ or $U=129,490$) in the two datasets. In addition, we only predict the $\hat{r}_{ij}$ for the $r_{ij} \neq 0$. Obviously, a lower $MAE$ denotes a higher prediction accuracy, e.g., $MAE=0$ means the prediction is totally correct because the prediction ratings equal to the real ratings, but no privacy guarantee against $k$NN attack.

\subsection{Experimental Results}
In this section, we show the accuracy performance from different perspectives of four main neighbourhood-based CF methods, i.e., $k$ Nearest Neighbour ($k$NN), naive Probabilistic Neighbour Selection (nPNS) \cite{ADAMOPOULOS2014}, Private Neighbour CF (PNCF) \cite{ZHU2014} and our method, Partitioned Probabilistic Neighbour Selection (PPNS). Due to the similarity metric (Cosine-based similarity, Equation \eqref{COS}) used in this paper, in the second half of a candidate list, a large number of candidates' similarity will be zero which is useless for prediction. So in the experiments, we set the upper bound of $\beta$ as $U/2k$ (user-based prediction) or $I/2k$ (item-based prediction).

\subsubsection{Accuracy performance with no attacking}
\begin{figure}[!h]
\centering
\includegraphics[width=3.65in]{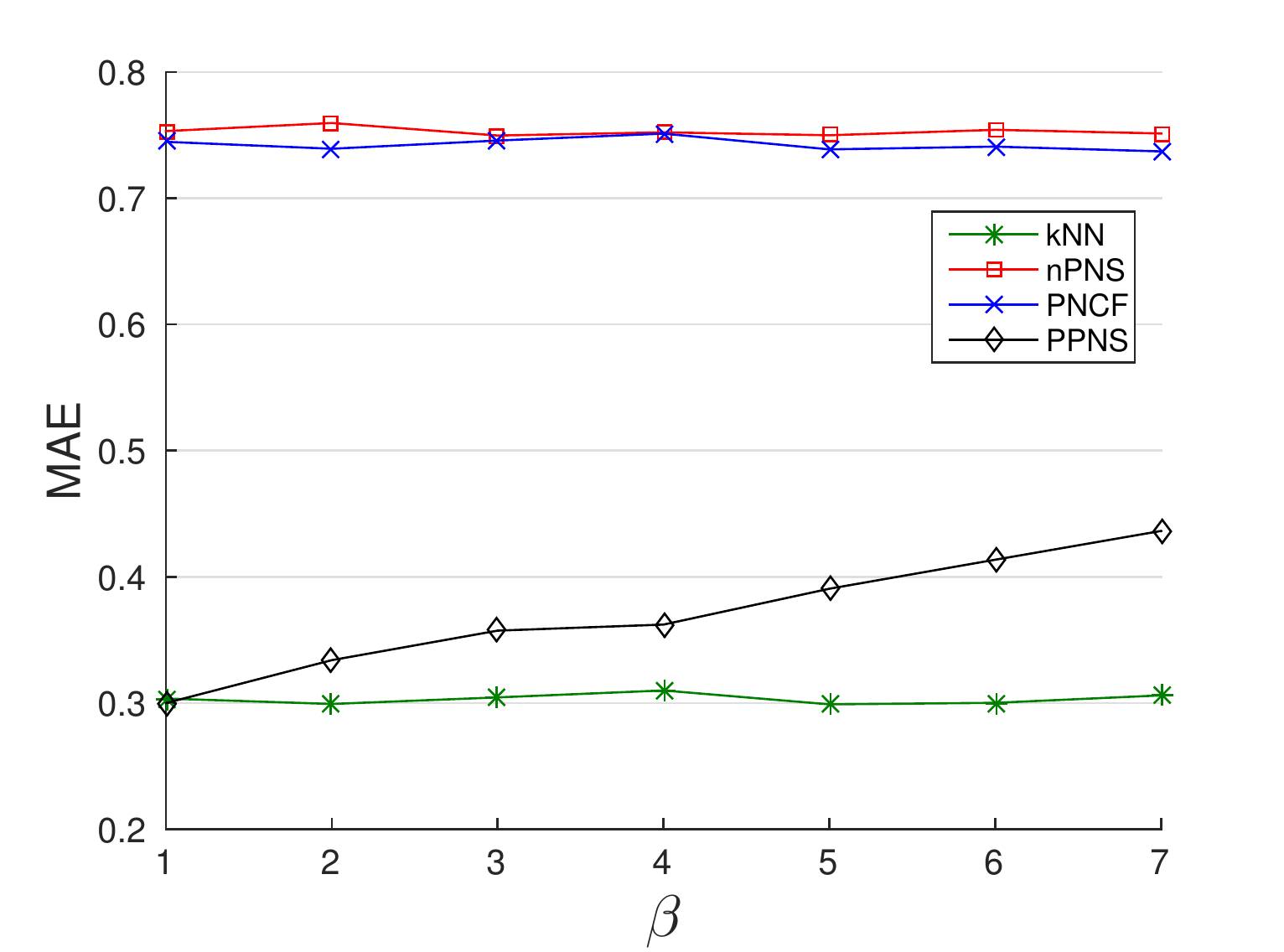}
\caption{Item-based prediction accuracy on MovieLens ($\epsilon=1$, $k=100$)}
\label{fig:N1item}
\end{figure}

\begin{figure}[!h]
\centering
\includegraphics[width=3.65in]{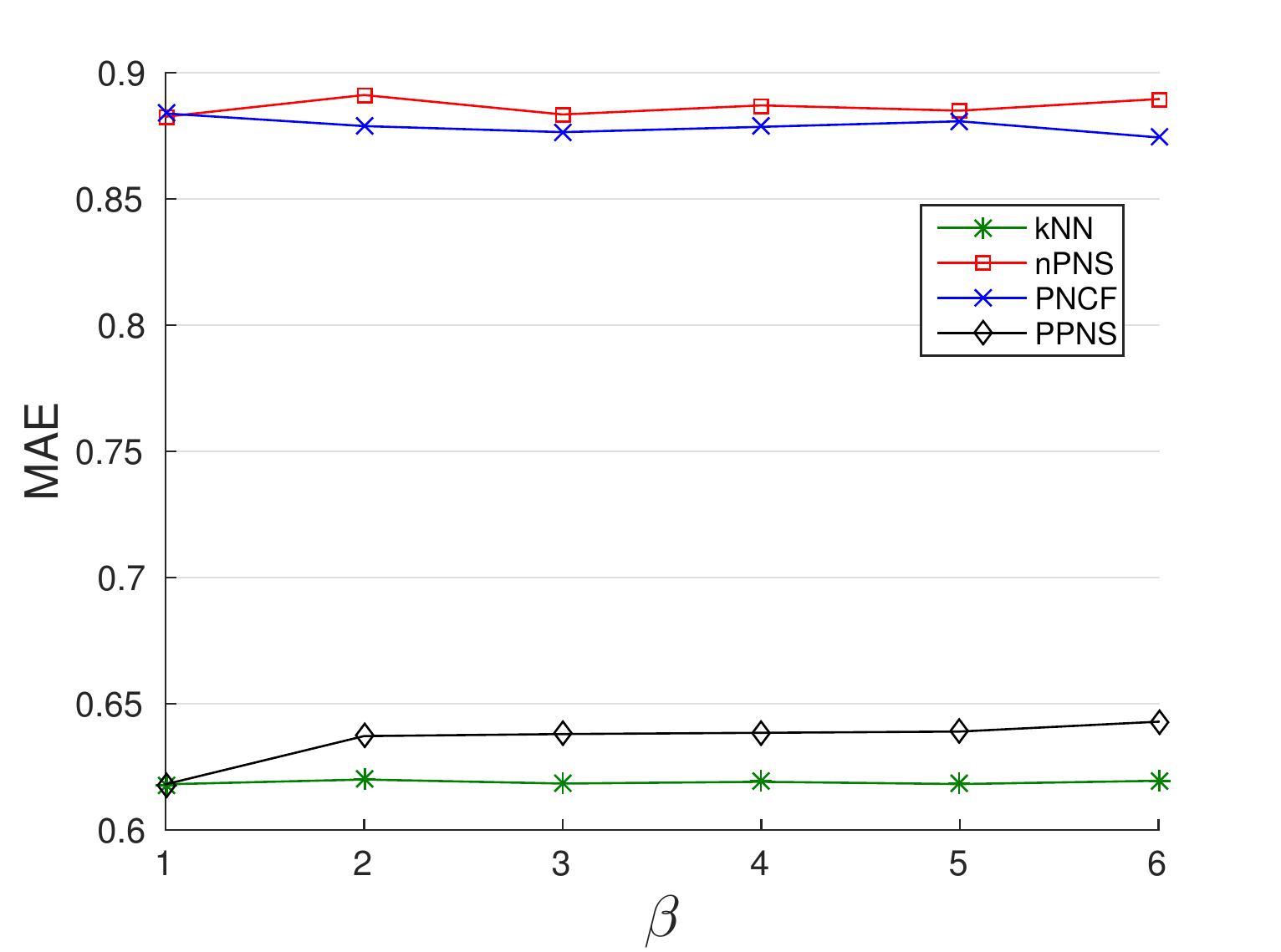}
\caption{User-based prediction accuracy on MovieLens ($\epsilon=1$, $k=100$)}
\label{fig:N1user}
\end{figure}

\begin{figure}[!h]
\centering
\includegraphics[width=3.65in]{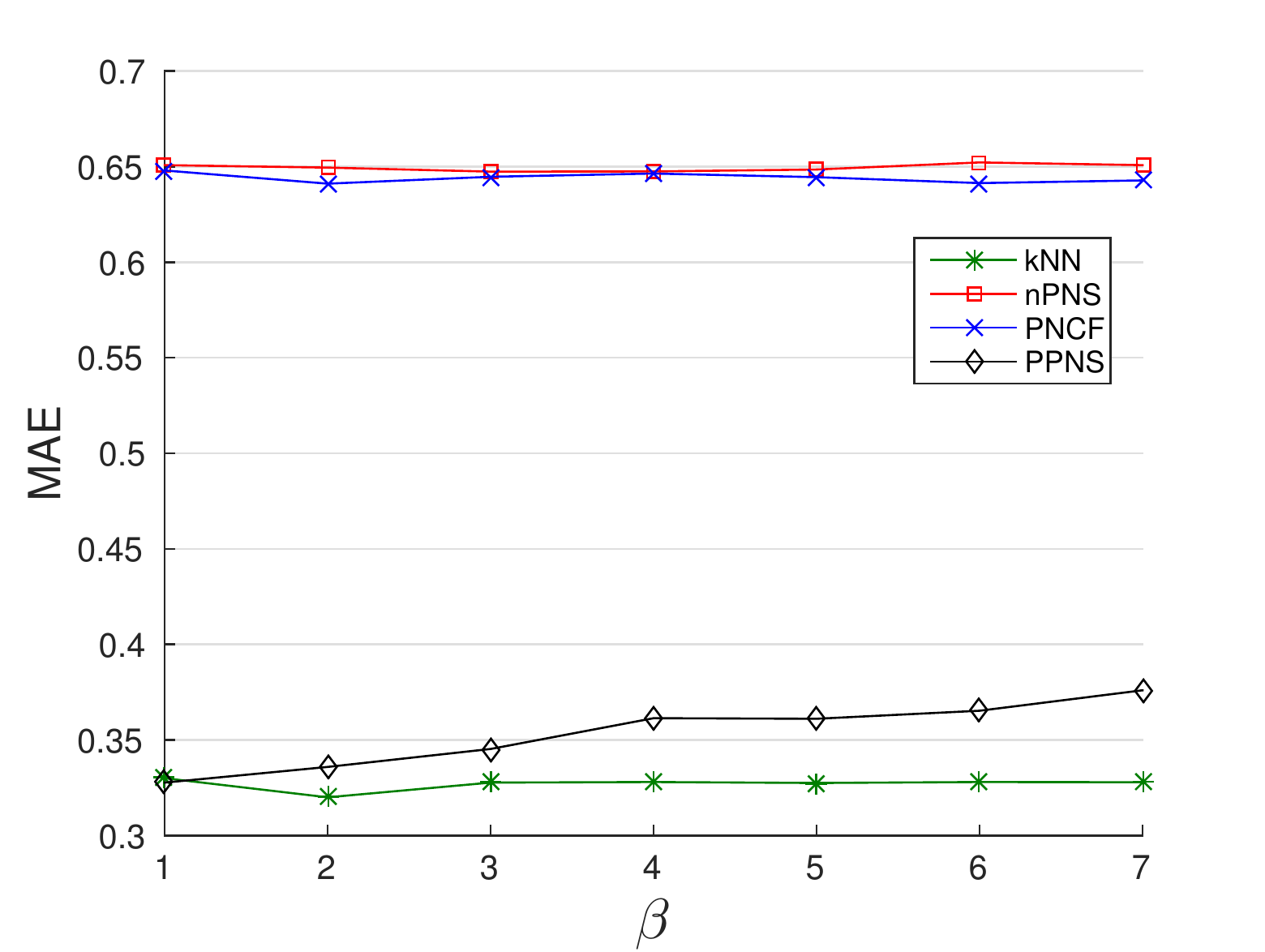}
\caption{User-based prediction accuracy on Douban film ($\epsilon=1$, $k=100$)}
\label{fig:N2user}
\end{figure}

We design three experiments (Figure \ref{fig:N1item} - Figure \ref{fig:N2user}) to examine the user-based and item-based CF prediction accuracy on MovieLens dataset and Douban film dataset. As seen in Figure \ref{fig:N1item} to Figure \ref{fig:N2user}, we notice that our privacy preserving method (PPNS) achieves much better accuracy performance than the two global methods (nPNS and PNCF) in both the two datasets on both user-based and item-based CF. Moreover, as a trade-off between the prediction accuracy and system security in PPNS, a greater security metric $\beta$ results in a greater $MAE$ which means a worse prediction accuracy. Specifically, when $\beta=1$, PPNS achieves the same prediction accuracy with the $k$NN method which is regarded as the baseline neighbourhood-based CF recommendation method in this paper.


\subsubsection{Accuracy performance against $k$NN attack}
To examine the accuracy performance of the four methods against $k$NN attack with the same security guarantee, we introduce a fixed security metric $\beta$ to the three privacy preserving CF algorithms (nPNS, PNCF, PPNS). That is, we randomly select $k$ neighbours from the $\beta k$ nearest candidates with weighted sampling in nPNS; we calculate $\lambda$ as $sim_k-sim_{\beta k}$ in PNCF; and we select the $k$ neighbours across the top $\beta$ partitions by Algorithm \ref{ALG} in PPNS. The experiments are run on user-based CF because $k$NN attack is a user-based attacking.

\begin{figure}[!h]
\centering
\includegraphics[width=3.65in]{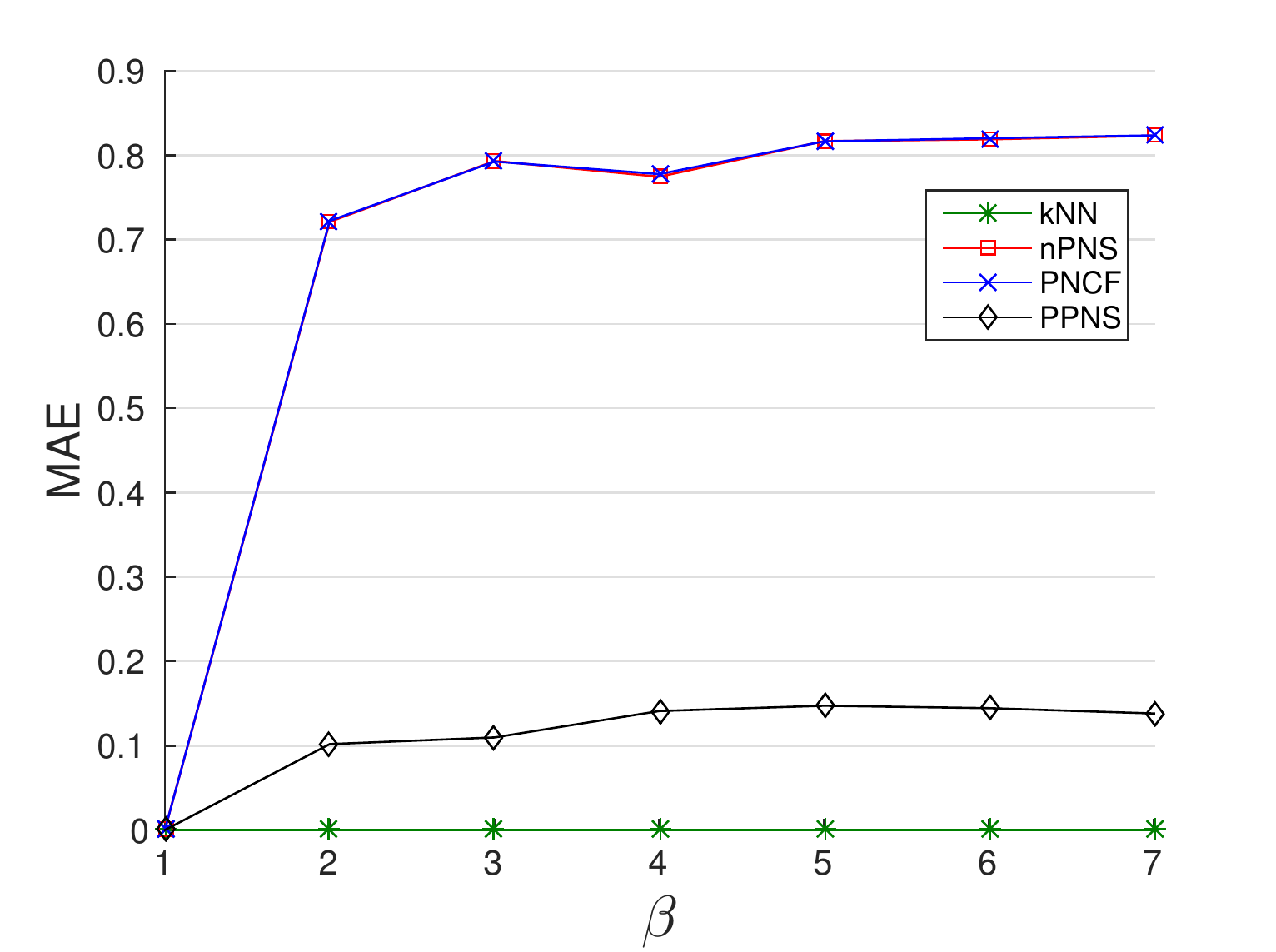}
\caption{Prediction accuracy on MovieLens against $k$NN attack ($\epsilon=1$, $k=50$, $m=8$)}
\label{fig:A1user}
\end{figure}

Figure \ref{fig:A1user} shows that to ensure the same security guarantee against $k$NN attack, PPNS performs much better on the prediction accuracy than the other privacy preserving CF methods (nPNS and PNCF). Moreover, the $MAE$ performance of the $k$NN method indicates that $k$NN CF does not provide any security guarantee against $k$NN attack. Additionally, as we regard $\beta$ as security metric, we observe that we achieve a trade-off between accuracy and security, because the greater $\beta$ yields a greater $MAE$ which denotes less prediction accuracy.

\begin{figure}[!h]
\centering
\includegraphics[width=3.65in]{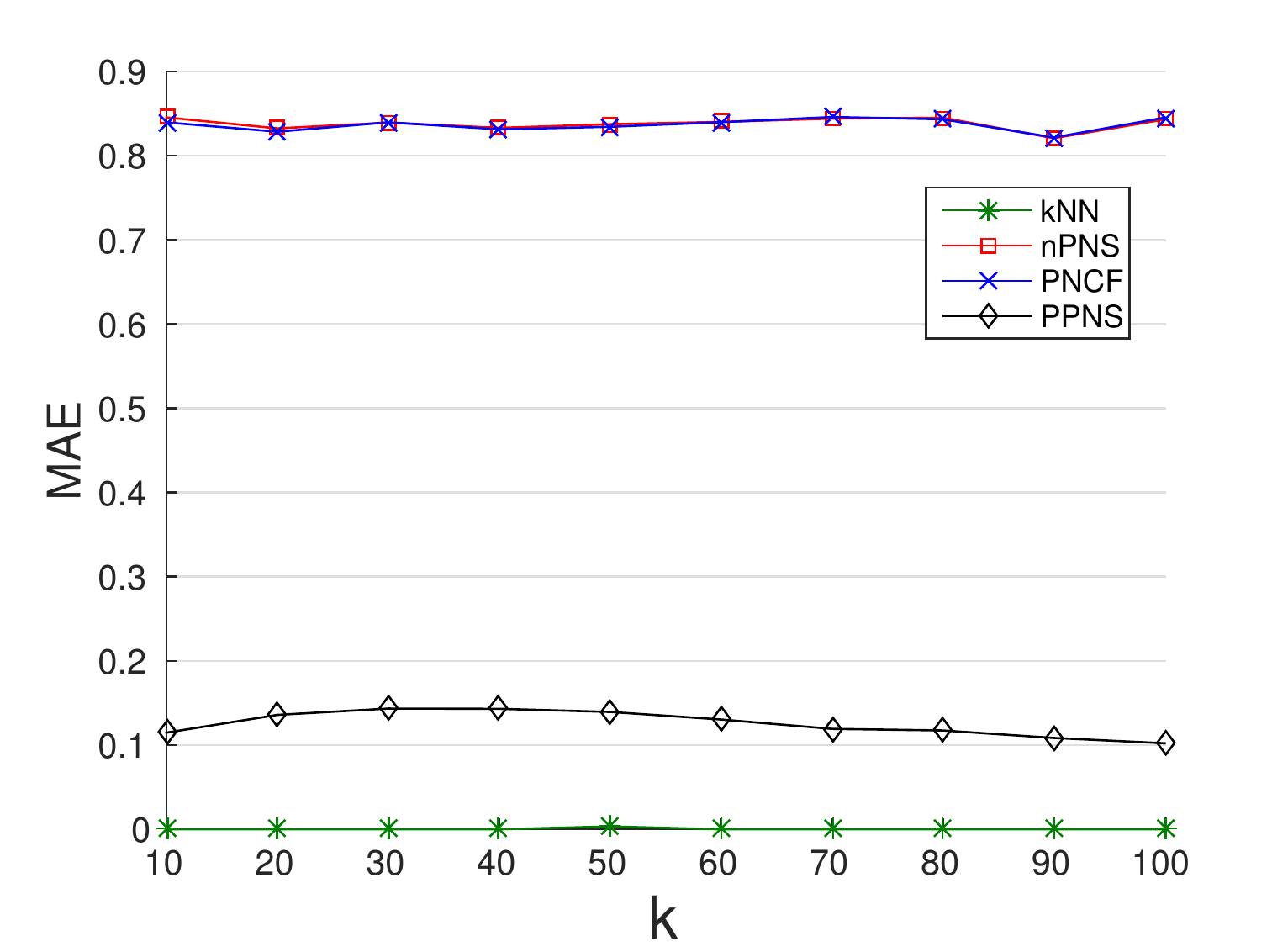}
\caption{Impacts of $k$ on prediction accuracy against $k$NN attack on MovieLens ($\epsilon=1$, $m=8$, $\beta=7$)}
\label{fig:k}
\end{figure}

\newpage
Figure \ref{fig:k} demonstrates the impacts of recommendation parameter $k$ on the prediction accuracy. We examine the value of $k$ from 10 to 100, which is a popular range for the recommendation parameter $k$. From Figure \ref{fig:k}, we can see that a larger size of neighbour set (or the size of partition in PPNS) denotes the better prediction accuracy of PPNS method against $k$NN attack.

\begin{figure}[!h]
\centering
\includegraphics[width=3.65in]{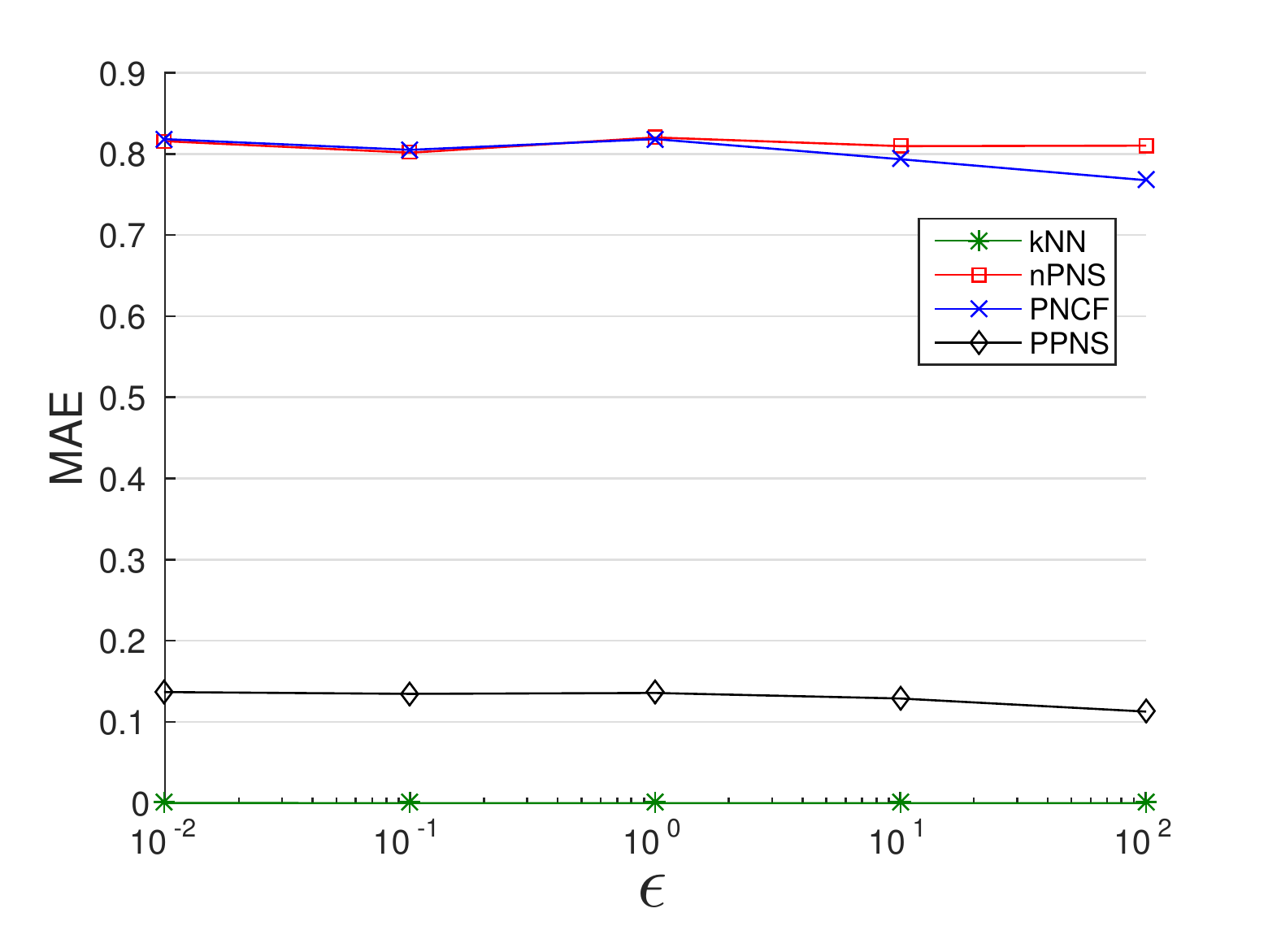}
\caption{Impacts of $\epsilon$ on prediction accuracy against $k$NN attack on MovieLens ($k=50$, $m=8$, $\beta=7$)}
\label{fig:epsilon}
\end{figure}

Figure \ref{fig:epsilon} illustrates the impacts of differential privacy budge $\epsilon$ on the prediction accuracy. It is observed that as $\epsilon$ increases, the $MAE$ performance improves in the two differential privacy methods (PNCF and PPNS). So to achieve a better prediction accuracy, it is suggested to set a greater $\epsilon$ against $k$NN attacks.

\begin{figure}[!h]
\centering
\includegraphics[width=3.65in]{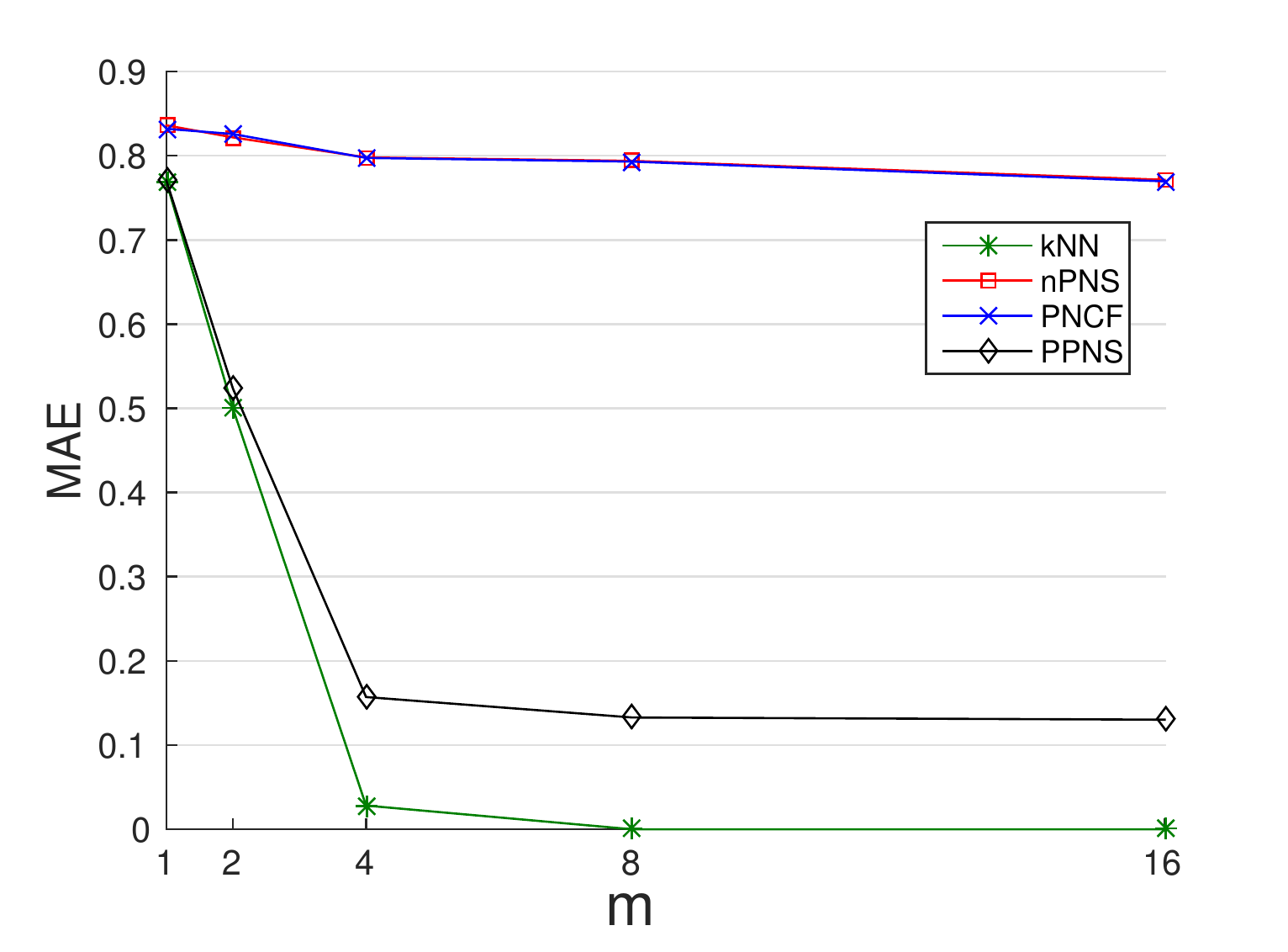}
\caption{Impacts of $m$ on prediction accuracy against $k$NN attack on MovieLens ($\epsilon=1$, $k=50$, $\beta=7$)}
\label{fig:m}
\end{figure}

Figure \ref{fig:m} presents the impacts of attacking parameter $m$ on the prediction accuracy. we can note that to reveal a target customer's privacy by $k$NN attack, the attacker needs at least $2^3$ real ratings of the target customer as auxiliary information, since when $m \geq 8$, the $MAE$ of a non-privacy preserving CF ($k$NN) method is zero. When the attacker has more background knowledge, the prediction will be closer to the real ratings for all of the neighbourhood-based CF systems, but none of privacy preserving algorithms releases the customer's privacy.

\section{Conclusion}
\label{CON}
Recommender systems play an important role in Internet commerce since the first decade of 21st century. To protect customers' private information against $k$NN attack during the process of filtering, the existing privacy preserving neighbourhood-based CF recommendation methods \cite{MCSHERRY2009,ZHU2014,ADAMOPOULOS2014} introduced global noise into the covariance matrix and the process of neighbour selection. However, they neither ensure the prediction accuracy because of the global noise, nor guarantee an assured security enforcement before the collaborative filtering against $k$NN attack. To overcome the weaknesses of the current probabilistic methods, we propose a novel privacy preserving neighbourhood-based CF method, Partitioned Probabilistic Neighbour Selection, to ensure a required security while achieving the optimal prediction accuracy against $k$NN attack. The theoretical and experimental analysis show that achieving the same security guarantee against $k$NN attack, our method ensures the optimal performance of recommendation accuracy among the current randomised neighbourhood-based CF recommendation methods.

\bibliographystyle{abbrv}
\bibliography{zhigang} 

\end{document}